\documentclass[%
 twocolumn,
 superscriptaddress,
 amsmath,amssymb,
 aps,
]{revtex4-2}

\usepackage{graphicx}% Include figure files
\usepackage{dcolumn}% Align table columns on decimal point
\usepackage{bm}% bold math
\usepackage{mathtools, nccmath}
\usepackage{physics}
\usepackage{comment}
\usepackage{float}
\usepackage{xcolor}

\newcommand{\gs}{{\mathrm{gs}}}

\newcommand{\train}{{\mathrm{train}}}
\newcommand{\ts}{{\mathrm{ts}}}

\usepackage{hyperref}% add hypertext capabilities

\DeclarePairedDelimiter{\nint}\lfloor\rceil

\begin{document}

\preprint{APS/123-QED}

\title{Solving deep-learning density functional theory via variational autoencoders}

\author{Emanuele Costa}
\affiliation{Physics Division, School of Science and Technology, University of Camerino, Via Madonna delle Carceri 9, I-62032 Camerino (MC), Italy}
\affiliation{INFN Sezione di Perugia, Via A. Pascoli, I-06123 Perugia, Italy}

\author{Giuseppe Scriva}
\affiliation{Physics Division, School of Science and Technology, University of Camerino, Via Madonna delle Carceri 9, I-62032 Camerino (MC), Italy}
\affiliation{INFN Sezione di Perugia, Via A. Pascoli, I-06123 Perugia, Italy}

\author{Sebastiano Pilati}
\affiliation{Physics Division, School of Science and Technology, University of Camerino, Via Madonna delle Carceri 9, I-62032 Camerino (MC), Italy}
\affiliation{INFN Sezione di Perugia, Via A. Pascoli, I-06123 Perugia, Italy}

\begin{abstract}
In recent years, machine learning models, chiefly deep neural networks, have revealed suited to learn accurate energy-density functionals from data. However, problematic instabilities have been shown to occur in the search of ground-state density profiles via energy minimization. Indeed, any small noise can lead astray from  realistic profiles, causing the failure of the learned functional and, hence, strong violations of the variational property. In this article, we employ variational autoencoders to build a compressed, flexible, and regular representation of the ground-state density profiles of various quantum models. Performing energy minimization in this compressed space allows us to avoid both numerical instabilities and variational biases due to excessive constraints. Our tests are performed on one-dimensional single-particle models from the literature in the field and, notably, on a three-dimensional disordered potential. In all cases, the ground-state energies are estimated with errors below the chemical accuracy and the density profiles are accurately reproduced without numerical artifacts.
\end{abstract}


\maketitle

\section{Introduction}

Since a few decades, density functional theory (DFT) has arguably been  the most popular and effective simulation technique for solid-state systems and for chemical compounds~\cite{PhysRev.136.B864,doi:10.1063/1.4704546}. It allows scientists to predict the electronic properties at a feasible computational cost, in particular in its orbital-free implementation~\cite{Wang2002,doi:10.1142/8633}. However, the exact form of the universal energy-density functional is unknown, and the available approximations often fail in the presence of strong electron correlations~\cite{cohen2012challenges}.

In recent years, machine learning (ML) techniques have been introduced in the framework of DFT~\cite{PhysRevLett.108.253002}, addressing continuous-space systems~\cite{Brockherde2017,https://doi.org/10.1002/qua.25040,PhysRevA.100.022512}, as well as lattice~\cite{PhysRevB.99.075132,PhysRevLett.125.076402} and spin models~\cite{PhysRevB.108.125113}. The main goal is to learn from data more reliable energy-density functionals, potentially adequate also for strongly correlated systems. 
The envisioned strategy consists of training ML models, e.g., deep neural networks (NNs), exploiting datasets of ground-state energies and density profiles generated via an accurate but computationally expensive method. In principle, this would then allow one to determine the ground-state properties of novel system instances by simply minimizing the deep-learning (DL) functional, leading to a substantial reduction of computational cost.
Unfortunately, in actual implementations of energy minimization, severe problems have emerged, in particular when exploiting gradient-descent methods~\cite{Meyer2020,PhysRevLett.108.253002,https://doi.org/10.1002/qua.24937,PhysRevE.106.045309,snyder2013orbital,https://doi.org/10.1002/qua.25040}. Indeed, even minuscule inaccuracies of the functional get amplified. This leads to the formation of noisy density profiles that cannot be properly processed by the DL functionals, leading to large errors and violations of the variational property.
A few strategies to circumvent these instabilities have already been proposed. 
They focus mostly on reducing noise in the gradient via dimensionality reduction~\cite{https://doi.org/10.1002/qua.24937} or basis truncation~\cite{PhysRevB.94.245129}, on constraining the optimization, on training functionals with derivatives data~\cite{doi:10.1126/science.abj6511,2022,Meyer2020}, or on implementing tailored NN architectures that lead to more stable derivatives~\cite{PhysRevE.106.045309}. These studies addressed mostly low-dimensional single-particle models with random potentials, since these allow one to analyze the above instabilities while nimbly creating data for training and testing due to the affordable computational cost.

\begin{figure*}
\includegraphics[width=\textwidth]{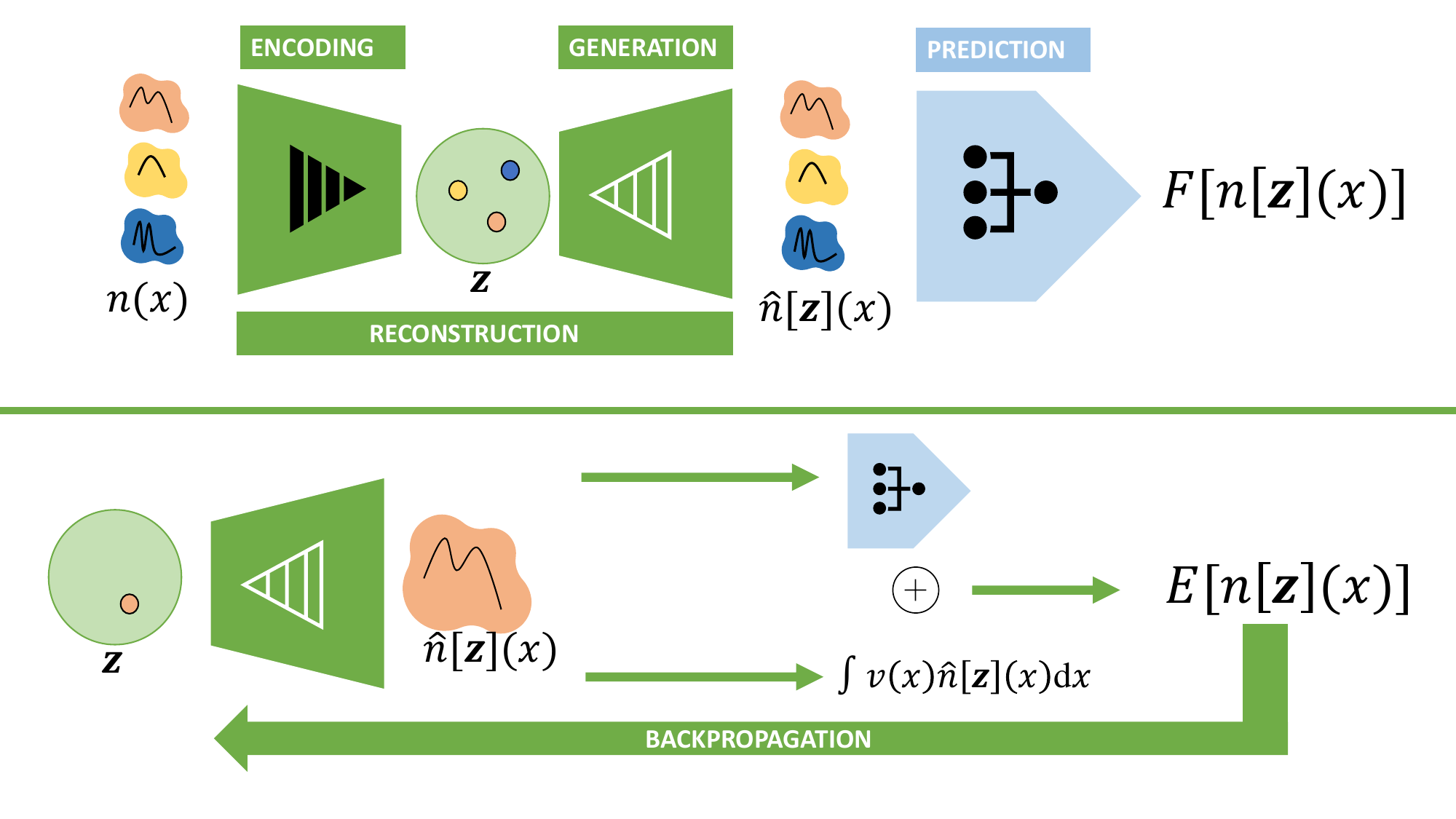}
\caption{\label{fig:figure1}Schematic representation of the $\beta-$VAE, of the DL functional, and of the combination of these architectures to perform a stable minimization of the DL functional. The optimization is performed in the latent space of the VAE such that only the manifold of actual ground state densities is explored.}
\end{figure*}

This article presents an alternative, particularly versatile, strategy to solve DL-DFTs. It exploits DL also to automatically guide the energy minimization, not only to learn the energy-density functional. Specifically, this strategy involves learning a compressed encoding of virtually all realistic profiles using a variational auto-encoder (VAE). This is achieved by training the VAE to accurately reproduce a dataset of exact ground-state densities. Then, this VAE is combined with a separate NN that maps the density to the corresponding ground-state energy. Finally, automatic differentiation is exploited to perform gradient-descent minimization in the encoded space of the VAE. 
The overall approach is schematically represented in Fig.~\ref{fig:figure1}.
As we numerically demonstrate, our strategy allows exploring virtually all realistic profiles, without  preventing the descent from reaching the actual ground state, and at the same time it avoids instabilities and violations of the variational property.

Our investigation focuses on three testbed models. The first two are one-dimensional (1D) single-particle models taken from the previous literature on this topic~\cite{PhysRevE.106.045309,Meyer2020}.
Notably, the third is a challenging three-dimensional (3D) potential formed via multiple Gaussian bumps with random centers and amplitudes.
In all cases, the VAE allows us to accurately solve the DL-DFT, reaching ground-state energies with errors below the chemical accuracy and with accurate density profiles, free of numerical artifacts.
To favor future investigations on DL-DFTs for 3D systems, we provide a large dataset of ground-state energies and densities of the 3D Gaussian model at the repository of Ref.~\cite{emanuele_2024_10814856}.

The rest of the article is organized as follows:
Sec.~\ref{sec:vae-dft} describes the DL-DFT approach, the NN used to learn the energy-density functional, and the VAE used to encode the density profiles. It also explains how to perform energy minimization in the encoded space of the VAE making use of automatic differentiation.
In Sec.~\ref{sec:models}, our three testbed models are described.
The results on density encoding and, chiefly, on solving DL-DFT via energy minimization are reported in Sec.~\ref{sec:results}.
In Sec.~\ref{sec:conclusion}, a summary of our main findings is reported.

\section{VAE for Density Functional Theory}
\label{sec:vae-dft}

We develop and test the DL-DFT method~\cite{https://doi.org/10.1002/qua.25040,PhysRevA.100.022512,2022,10.1088/2516-1075/ac572f} addressing single-particle Hamiltonians, defined as
\begin{equation}
\label{H}
    H=-\frac{\hbar^2}{2m}\nabla^2 + V(x),
\end{equation}
where $\hbar$ is the reduced Planck constant, $m$ is the particle mass, $V(x)$ is the external potential, and with $x$ we denote the particle coordinate. Our focus is on one-dimensional and on more challenging three-dimensional models. 
The chosen energy unit is $E_0=\frac{\hbar^2}{ma_0^2}$, where $a_0$ is the length unit. If $m$ is identified with the electron mass and $a_0$ with the Bohr radius, the energy unit corresponds to the Hartree energy.

The aim of DFT is to map the ground-state density profile, which in the single-particle case is computed as
\begin{equation}
    n_{\gs}(x)= |\psi_{\gs}(x)|^2,
\end{equation}
where $\psi_{\gs}(x)$ is the ground-state wave function, to the ground-state energy $e_{\gs}$.
The first Hohenberg-Kohn theorem guarantees the existence of this mapping~\cite{PhysRev.136.B864}.
In practice, it is convenient to separate the known potential energy contribution, seeking for the universal functional $f_{\gs}=e_{\gs}-\int \dd x V(x) n(x)$. In the single-particle scenario, this universal functional corresponds to the kinetic functional term~\footnote{In interacting systems, the unknown functional would also include correlation effects, while the mean-field contribution would be conveniently encoded in the so-called Hartree term.}.
The second Hohenberg-Kohn theorem defines the variational property for the energy functional $E[n]=F[n]+\int \dd x n(x) V(x)$:
\begin{equation}
    E[n] \ge e_{\gs},
\end{equation}
where the equality holds when $n(x)=n_{\gs}(x)$.
Many DFT studies introduce the Kohn-Sham formalism~\cite{PhysRev.140.A1133}, which provides one with a suitable approximation for the kinetic energy functional. This comes at the cost of introducing a set of orbitals, which are typically found via self-consistent iterations. Here, as in previous studies of DL-DFT approaches~\cite{PhysRevLett.108.253002,Meyer2020,PhysRevA.100.022512,2022,PhysRevE.106.045309,https://doi.org/10.1002/qua.24937,https://doi.org/10.1002/qua.25040,Brockherde2017}, we adopt the computationally cheaper orbital free approach, where the ground-state properties of the many-body system are obtained via minimization of the energy functional $E[n]$.
The long-term ambition is to obtain, via orbital-free DL approaches, comparable if not superior accuracy than in Kohn-Sham schemes at a reduced computational cost.

\subsection{Neural Networks} \label{sec:nn}

\subsubsection{Variational Autoencoder} \label{subsec:vae}
Variational autoencoders (VAEs)~\cite{Kingma2014,higgins2017betavae} are specific instances of autoencoders, typically used for dimensionality reduction and image generation. 
They are defined by two conditional probabilities. There first is the  encoder conditional probability, fixed by the density profile, and defined in the latent space
\begin{equation}
    p_{\phi}(\mathbf{z}|\bm{n}) = \mathcal{N}(\bm{\mu}_{\phi}[\bm{n}], \bm{\sigma}_{\phi}[\bm{n}]),
\end{equation}
where, in 1D, the discretized density $\bm{n}=\left(n(x_1),n(x_2),...,n(x_{N_g})\right)$ is defined over a uniformly spaced grid of $N_g$ points (the generalization to higher dimensions is straightforward). $\bm{\mu}_{\phi}$ and $\bm{\sigma}_{\phi}$ are outputs of a neural network defined in $\mathbb{R}^{l_d}$, and $\mathcal{N}(\bm{\mu},\bm{\sigma})$ indicates the multivariate normal distribution with mean $\bm{\mu}=\left( \mu_1,\mu_2,...,\mu_{l_d}\right)$ and standard deviation $\bm{\sigma}=\left(\sigma_1,\sigma_2,...,\sigma_{l_d}\right)$. 
The second conditional probability, namely, the decoder conditional probability, is fixed by the corresponding latent variable $\mathbf{z}$ and it  is defined in the density profile manifold as
\begin{equation}
    q_{\theta}(\bm{n}|\mathbf{z}) = \mathcal{N}(D_{\theta}[\mathbf{z}],\bm{1}),
\end{equation}
where $D_{\theta}[\mathbf{z}] \in \mathbb{R}^{N_g}$.
VAEs are designed for a two-fold goal, namely, accurately reconstructing the ground-state density profiles through the compressed latent space and generating novel realistic density profiles from points sampled in the latent space.
This is achieved by appropriately regularizing the latent space. For this, loss function is defined as follows:
\begin{equation}
    L(\theta,\phi)=  L_{\mathrm{rec}} + \beta L_{\mathrm{reg}}.
    \label{final loss vae}
\end{equation}
$L_{\mathrm{rec}}$ is the reconstruction loss
\begin{equation}
    L_{\mathrm{rec}}=\sum^{N_{\mathrm{train}}}_{r=1}\mathbb{E}_{p_{\phi}(\mathbf{z}|\bm{n}^{(r)})} \log q_{\theta}(\bm{n}^{(r)}|\mathbf{z}),
\end{equation}
where $\mathbb{E}$ indicates the expectation value with respect to a probability distribution and $N_{\mathrm{train}}$ is the number of training instances. $L_{\mathrm{reg}}$ is the regularization loss
\begin{equation}
    L_{\mathrm{reg}}=\sum^{N_{\mathrm{train}}}_{r=1} \mathrm{KL}(p_{\phi}\left(\mathbf{z}|\bm{n}^{(r)})\,||\,p(\mathbf{z})\right),
\end{equation}
where $\mathrm{KL}(p||q)$ denotes the Kullback-Leibler divergence between the probability distributions $p$ and $q$, and $p(\mathbf{z})$ is the prior probability, which corresponds to the standard normal distribution $\mathcal{N}(\mathbf{0},\mathbf{1})$. 

For $\beta=1$ the total loss corresponds to the evidence lower bound~\cite{Kingma2014, higgins2017betavae, qian-cheung-2019-enhancing}. In general, $\beta$ can be treated  as a hyperparameter; it controls the interplay between the regularization of the latent space and the reconstruction accuracy~\cite{9244048,higgins2017betavae}.
The regularization loss forces the conditional distributions $p_{\phi}(\mathbf{z}|\bm{n})$ to resemble the standard normal. In this way, the components of the latent variables can be made less entangled~\cite{higgins2017betavae, burgess2018understanding,mathieu2019disentangling} and the overlap between distributions corresponding to different inputs can be increased. These effects contribute to a more dense and regular structure of the latent space, but this comes at the cost of lower reconstruction accuracy. 
As we show in Sec.~\ref{subsec:grdient-desc}, $\beta$ plays indeed an important role in the convergence and in the stability of the gradient-descent minimization of the DL-DFT.
Another important hyperparameter is the latent space dimension $l_d$, which determines whether all relevant information of the density profiles can be extracted. This hyperparameter must be tuned depending on the complexity of the problem, specifically, on the variability of the density profiles corresponding to different Hamiltonian instances. We analyze this effect considering different disordered potentials which lead, depending on the random parameters of the Hamiltonian, either to rather consistent or to quite variable density profiles.

The encoder network is composed of a series of convolutional blocks. Each block is made of a convolutional layer, a smooth activation function called Softplus~\cite{7280459}, an average pooling operation, and a batch normalization~\cite{ioffe2015batch}. At the end of the series of convolutional blocks, two dense heads made by three dense layers, with respectively $100, 50$, and $l_d$ hidden neurons with the Softplus activation function, process the output of the convolutional blocks and return $\bm{\sigma}_{\phi}[\bm{n}]$ and $\bm{\mu}_{\phi}[\bm{n}]$.

The decoder network is composed symmetrically to the encoder. The latent variable is processed by a linear operation that returns the input of the forward convolutional blocks. The input shape is fixed such that the output of the series of block convolutions has the same shape as the input of the encoder. The convolutional blocks are composed of transpose convolutional layers, a Softplus activation function, and batch normalization. The last block features the identity activation function. 
To take into account the constraints of the density profile, namely, normalization and positivity, the output is processed by a sigmoid layer~\cite{10.1007/3-540-59497-3_175}, followed by a normalization operation. The normalization is performed by applying the numerical integration via sum rule $\int \dd x f(x) \rightarrow \Delta x \sum_{i} f(x_{i})$, where the generalization to higher dimension is straightforward.
The same rule is applied to all numerical integrations, and it is verified that $N_g$ is large enough to suppress the effect of the discretization error below the chosen target.

The training process is performed using the reparameterization trick described in Ref.~\cite{Kingma2014}. 
The training is performed using adaptive stochastic gradient descent (ADAM)~\cite{kingma2014adam} for $1200$ epochs, batch size $100$ and a learning rate equal to $10^{-4}$. The split training/validation is $80 \% / 20 \%$ for all the datasets considered.
Tab.~\ref{tab:table1} resumes all the hyperparameters adopted for the training.
\begin{table}
\caption{\label{tab:table1}%
Hyperparameters for VAE. 
}
\begin{ruledtabular}
\begin{tabular}{lccr}
&\textrm{1D Gaussian}&\textrm{1D speckle}&\textrm{3D Gaussian}\\
\colrule
\textrm{Epochs} & 1200 & 1200 & 1200\\
\textrm{Learning rate} & $10^{-4}$ & $10^{-4}$ & $10^{-4}$\\
\textrm{Batch} & 100 & 100 & 100\\
\textrm{Latent space} & 4,8 & 16 & 32\\
\textrm{Conv. channels} & 60 & 60 & 60, 120, 180\\
\textrm{Conv. layers} & 5 & 5 & 3\\
\text{Neurons (dense lay.)} & 100, 50, $l_d$ & 100, 50, $l_d$ & 100, 50, $l_d$\\
\textrm{Kernel} & 13 & 13 & 3\\
\textrm{Pooling} & 2 & 2 & 2\\
\text{Optimizer} & ADAM & ADAM & ADAM \\
\text{Act. func.} & Softplus & Softplus & Softplus \\
\end{tabular}
\end{ruledtabular}
\end{table}

\subsubsection{DL-functional} \label{subsec:dl-func}
The DL-functional $\Tilde{F}_{\omega} [n_{\gs,k}]$ is trained to map the density profile  $n(x)$ to the corresponding universal functional value $f_{\gs}$. We consider a supervised learning approach by using a dataset $\{ n_{\gs,k}, f_{\gs,k} \}$, where the index $k$ labels Hamiltonian instances. The network parameters $\omega$ are optimized by minimizing the mean squared error loss function
\begin{equation}
    \mathcal{L}(\omega)= \frac{1}{N_{\train}}\sum_{k=1}^{N_{\train}}| f_{\gs,k}- \Tilde{F}_{\omega} [n_{\gs,k}]|^2.
\end{equation}

The neural network is composed of a series of convolutional blocks, each including a convolutional layer, a Softplus activation operation~\cite{7280459}, and an average pooling. The output of the convolutional part is processed by a linear projector that outputs the universal functional value.
As for the VAE, we adopt again adaptive ADAM algorithm for $1200$ epochs, batch size $100$ and a learning rate equal to $10^{-4}$. The split training/validation is $80 \% / 20 \%$ for all the datasets considered.
Tab.~\ref{tab:table2} resumes all  training hyperparameters.
\begin{table}
\caption{\label{tab:table2}%
Hyperparameters for DL-functional. 
}
\begin{ruledtabular}
\begin{tabular}{lccr}
&\textrm{1D Gaussian}&\textrm{1D speckle}&\textrm{3D Gaussian}\\
\colrule
\textrm{Epochs} & 1200 & 1200 & 1200\\
\textrm{Learning rate} & $10^{-4}$ & $10^{-4}$ & $10^{-4}$\\
\textrm{Batch} & 100 & 100 & 100\\
\textrm{Conv. channels} & 60 & 60 & 60\\
\textrm{Conv. layers} & 5 & 5 & 4\\
\textrm{Kernel} & 13 & 13 & 3\\
\textrm{Pooling} & 2 & 2 & 2\\
\text{Optimizer} & ADAM & ADAM & ADAM\\
\text{Act. func.} & Softplus & Softplus & Softplus\\

\end{tabular}
\end{ruledtabular}
\end{table}

The neural networks, the trainings and the gradient descent method are implemented using the PyTorch library~\cite{NEURIPS2019_9015} and executed on a NVIDIA RTX A6000 GPU. 

\subsection{Gradient Descent Optimization}
\label{subsec:grdient-desc}

In the orbital free DL-DFT framework, the ground-state energy and density of novel Hamiltonian instances are determined by minimizing the DL functional using the gradient descent algorithm. The latter is defined by the following iterative step:
\begin{equation}
    n_{t+1}(x)= n_{t}(x) - \eta \left( \frac{\delta \Tilde{F}_{\omega}[n_t]}{\delta n(x)} +V(x) - \mu_{t} \right),
    \label{n gradient descent}
\end{equation}
where $\eta>0$ is the chosen learning rate, the integer $t=0,1,\dots,t_{\mathrm{max}}$ labels the steps, and the coefficient $\mu_{t}$ can be adapted, if needed, to ensure the normalization condition:
\begin{equation}
    \int \dd x \, n_t(x) = 1.
\end{equation}

Unfortunately, as discussed in several previous studies~\cite{PhysRevLett.108.253002,https://doi.org/10.1002/qua.24937,https://doi.org/10.1002/qua.25040,Meyer2020,PhysRevE.106.045309}, this minimization often fails in the presence of even minimal inaccuracies of the DL-functional derivative $\delta \Tilde{F}_{\omega}[n]/\delta n$. 
 Indeed, these inaccuracies lead the descent towards nonphysical density profiles, often causing strong violation of the variational property. Some methods have already been implemented to restore stability. They are based, e.g., on the regularization of the functional derivative by linear and non-linear principal component analysis~\cite{https://doi.org/10.1002/qua.24937,Meyer2020}, or on tailored NNs that reduces the noise of the functional derivative via an average operation performed on the hidden channels~\cite{PhysRevE.106.045309}.
In this article, an alternative strategy is introduced. The instability is avoided by performing the gradient-descent minimization within the latent manifold generated by a VAE, which is trained to reproduce virtually all realistic density profiles. 
The ground state energy is hence expressed as an effective functional of the latent variable $\mathbf{z}$ via the combination of the decoder NN that returns the corresponding density profile $\hat{n}[\mathbf{z}](x)=D_{\theta}[\mathbf{z}]$, and the DL-DFT functional:
\begin{equation}
    E[\mathbf{z}]= \Tilde{F}_{\omega}[\hat{n}[\mathbf{z}](x)] + \int \dd x \, v(x) \hat{n}[\mathbf{z}](x).
\end{equation}
This functional has a minimum in the latent configuration corresponding to the ground-state density profile
\begin{equation}
    \frac{\delta E[\mathbf{z}_{\gs}]}{\delta \mathbf{z}}=\int \dd x \left( \frac{\delta \Tilde{F}_{\omega}[n_{\gs}]}{\delta n_{\gs}(x)}+V(x) \right) \frac{\delta \hat{n}[\mathbf{z}_{\gs}](x)}{\delta \mathbf{z}}=0.
\end{equation}
Clearly, the actual ground state is reached only if it can be generated by decoding one of the points of the latent space. We will show that flexible enough VAEs can be easily implemented, while also avoiding unphysical artifacts that cause instabilities.
With the above rearrangement, the gradient-descent algorithm is written as:
\begin{equation}
    \mathbf{z}_{t+1}= \mathbf{z}_{t} - \eta \frac{\delta E[\mathbf{z}_t]}{\delta \mathbf{z}_t},
\label{gd with z}
\end{equation}
where $\eta>0$ is the learning rate related to the latent space. Eq.~\eqref{gd with z} does not contain the adaptive coefficient $\mu_t$ because the normalization constraint, as well as the positivity constraint, is already embodied in the decoder.
The value of $\eta$ is set also depending on the value of $\beta$, since the regularization loss  affects the norm of the latent variables. In our study $\eta$ ranges from $10^{-2}$ to $10^{6}$.
The number of gradient-descent steps ranges from $t_{\mathrm{max}}=9500$ to $t_{\mathrm{max}}=30000$, depending on the learning rate.
Suitable choices for the initial density profile $n_0(x)$ are the average density profile of the training dataset or one randomly chosen configuration of the same dataset.
Once the minimization is converged to a latent space point $\mathbf{z}_{\min}$, the ground-state energy and density profile can be estimated as $e_{\min}=E[\mathbf{z}_{\min}]$ and $n_{\min}(x)=\hat{n}[\mathbf{z}_{\min}](x)$, respectively.

\section{Testbed models and training Dataset} \label{sec:models}
In the following, we describe the three testbed models considered in this article.

\subsection{1D Gaussian barrier potential} \label{subsec:Gaussian1D}
The first testbed model is a 1D single particle Hamiltonian with a barrier formed by three Gaussians. It was originally introduced in Ref.~\cite{Meyer2020} to study ML-DFTs. The potential is defined as:
\begin{equation}
    V(x)=\sum_{i=1}^3 a_i \exp \left( -\frac{(x-b_i)^2}{2 c^2_i}\right),
\end{equation}
with $a_i,b_i,c_i$ randomly sampled from the uniform distribution in the intervals $[1E_0,10E_0]$, 
$[0.4a_0,0.6a_0]$, $[0.03a_0,0.1a_0]$. For this system, we consider a box of size $L=a_0$ and a grid of $N_g=256$ points, with hard-wall boundary conditions. 
For this testbed model, the density profiles corresponding to different random parameters display small variations. This is due to the chosen boundary conditions and to the selected ranges of  parameters. 
The training and testing datasets are obtained via exact diagonalization of the Hamiltonian matrix obtained via a nine-point finite-difference discretization. It includes $N_{\train}=15000$ instances.
For the test dataset, we use parameters from Ref.~\cite{Meyer2020}.

\subsection{1D speckle potential}
The second testbed is a 1D Hamiltonian with an external potential describing a random optical field. It was previously used in the framework of DL-DFT in Ref.~\cite{PhysRevE.106.045309}. This model represents the effect of disordered potentials used in cold-atom experiments. The potential can be numerically created from a  Gaussian random complex field, via the Fourier filter described in Refs.~\cite{goodman2007speckle,PhysRevA.73.013606}. 
At any point $x$, the intensity of the potential $V$ follows the probability distribution
\begin{equation}
    P(V)=\exp(-\frac{V}{V_0})
\end{equation}
where $V_0 \ge 0$ is the average intensity and $V \ge0$. The size of the speckle grains determines the characteristic disorder correlation length, which we set to coincide with the unit length $a_0$. Specifically, this length scale determines the extent of the two-point autocorrelation function, which reads:
\begin{equation}
    \frac{\overline{V(x' +x)V(x')}}{V^{2}_0} -1=\frac{ \sin(\pi x/ a_0)^2}{(\pi x / a_0)^2},
\end{equation}
where the bar indicates the average over a large ensemble of disorder realizations or, equivalently, over space in a sufficiently large box.
For the training and testing datasets, we fix the disorder strength at the intermediate value $V_0=0.25 E_0$, with the system size $L=14 a_0$, and a grid including $N_g=256$ points. Periodic boundary conditions are adopted.
Again, the ground-state energies and density profiles are determined via an eleven-point finite-difference method~\cite{PhysRevA.95.013613}.
The number of instances for the training dataset is about $N_{\train}=120 000$ and it is available at the repository Ref.~\cite{costa_emanuele_2022_6504567}.
Due to the large system size compared to the disorder correlation length, combined with the choice of periodic boundary conditions, different instances of the speckle potential typically lead to quite different density profiles. Hence, this potential represents a complementary testbed compared to the 1D Gaussian model.

\subsection{3D Gaussian potential}
\label{subsec:gaussian3D}
The third model is a 3D random potential, which is introduced here as a novel challenging testbed for DL-DFT. It allows the creation of  substantially different ground-state density profiles. The potential satisfies periodic boundary conditions in a 3D box of size $L=a_0$.
It is defined considering $N_{\mathrm{gauss}}=10$ scattering centers, with random positions ${\mathbf g}_i=(g_{xi},g_{yi},g_{zi})$, where $i=1,\dots,N_{\mathrm{gauss}}$, and the Cartesian coordinates $g_{\gamma i}\in[0,L]$, with $\gamma=x,y,z$, are sampled from a uniform random distribution in the box, namely, $g_{\gamma i} \sim  \mathrm{Uniform}([0,L])$. The effect of each scatterer is described as a Gaussian multiplied by a unique uniform random amplitude $A_i\sim \mathrm{Uniform}([0,2])$, but featuring the same standard deviation $\sigma$. 
Hence, the potential at position ${\mathbf r}=(x,y,z)$ is computed as:
\begin{equation}
V\left( {\mathbf r} \right) = 
\frac{E_0}{N_{\mathrm{gauss}}}\sum_{i=1}^{N_{\mathrm{gauss}}}
\frac{A_i}{(2\pi)^{3/2}(\sigma/a_0)^3}\exp\left[
- \frac{ \lVert {\mathbf \Delta r}_i \rVert^2 } {2\sigma^2}
\right];
\end{equation}
here, the distance from the $i$-th scatterer ${\mathbf \Delta r}_i=(\Delta x_i,\Delta y_i,\Delta z_i)$ is computed adopting the minimum image convention, namely, 
$\Delta \gamma_i = \delta \gamma_i - L \nint{\delta \gamma_i / L}$, where $\nint{\;}$ denotes the nearest-integer function, and $\delta \gamma_i = \gamma - g_{\gamma i}$. This allows complying with periodic boundary conditions. 
We set $\sigma=a_0/6$. With this choice, periodic images of scatterers beyond the nearest one are irrelevant.
The training dataset has $N_{\train}=36 000$ instances.
The ground-state properties are determined via an eleven-point finite difference formula for the 3D Laplacian~\cite{PhysRevA.95.013613} with a grid of $N_g=18$ points per direction.

\section{Results}
\label{sec:results}
\begin{figure}
\includegraphics[width=\columnwidth]{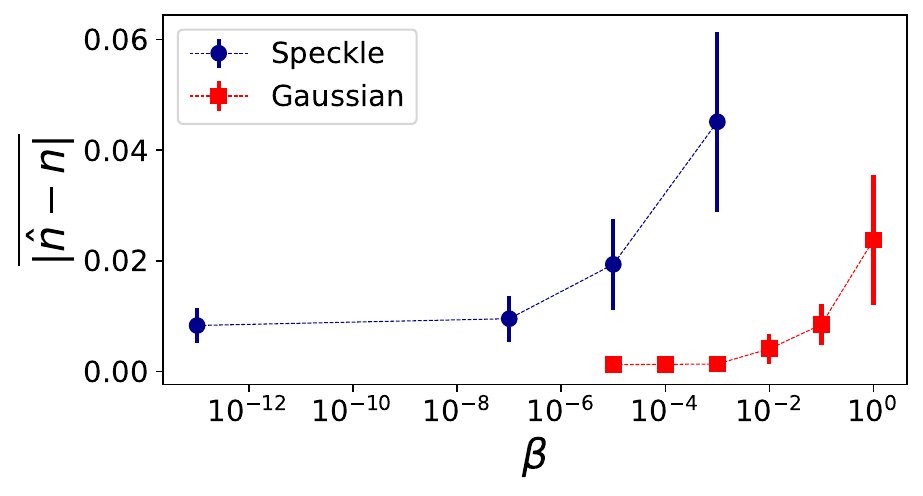}
\caption{\label{fig:reconstruction-vs-beta} Mean reconstruction error $\overline{|\bm{\hat{n}}-\bm{n}|}$ as a function of the (adimensional) regularization parameter $\beta$ for the 1D speckle potential (blue circles) and the 1D Gaussian potential (red squares). 
 The adimensional error measure $\overline{|\bm{\hat{n}}-\bm{n}|}$ is the integrated absolute density discrepancy, averaged over $N_{\ts}=100$ test instances. The error bar denote the standard deviation.
 In both cases, the reconstruction accuracy worsens as  $\beta$ increases.
 }
\end{figure}
\begin{figure}
\includegraphics[width=\columnwidth]{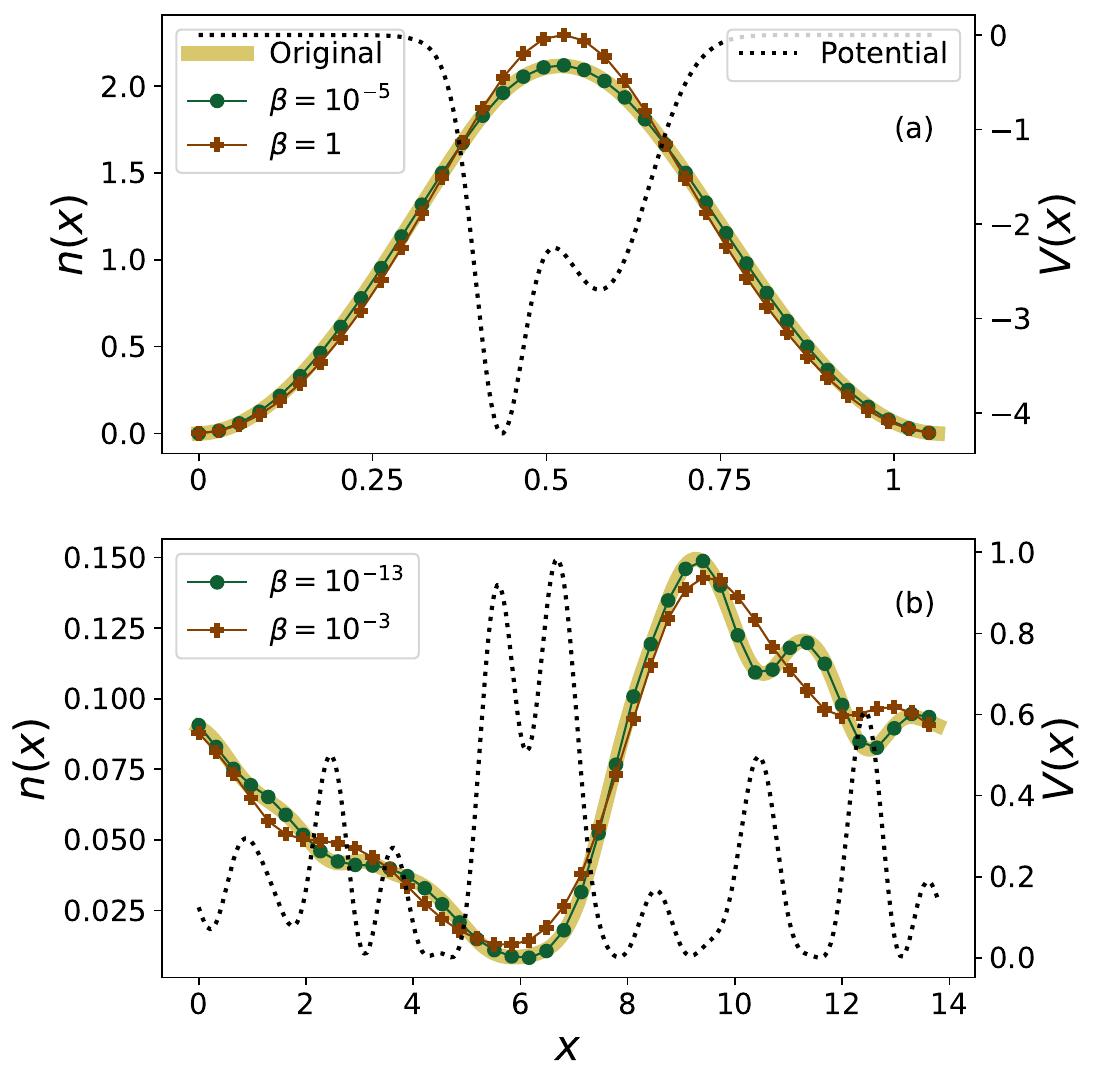}
\caption{\label{fig:reconstruction-samples}Illustrative examples of reconstructed density profiles compared to the ground-truth data $n(x)$ (thick yellow curve), for the 1D Gaussian potential (a) and for the 1D speckle potential (b). Different values of the hyperparameter $\beta$ are considered. The length and density units are $a_0$ and $1/a_0$, respectively. The corresponding external potentials $V(x)$ are also shown (dotted curves), referred to the right vertical axis in units of $E_0$.}
\end{figure}
\begin{figure*}
\includegraphics[width=\textwidth]{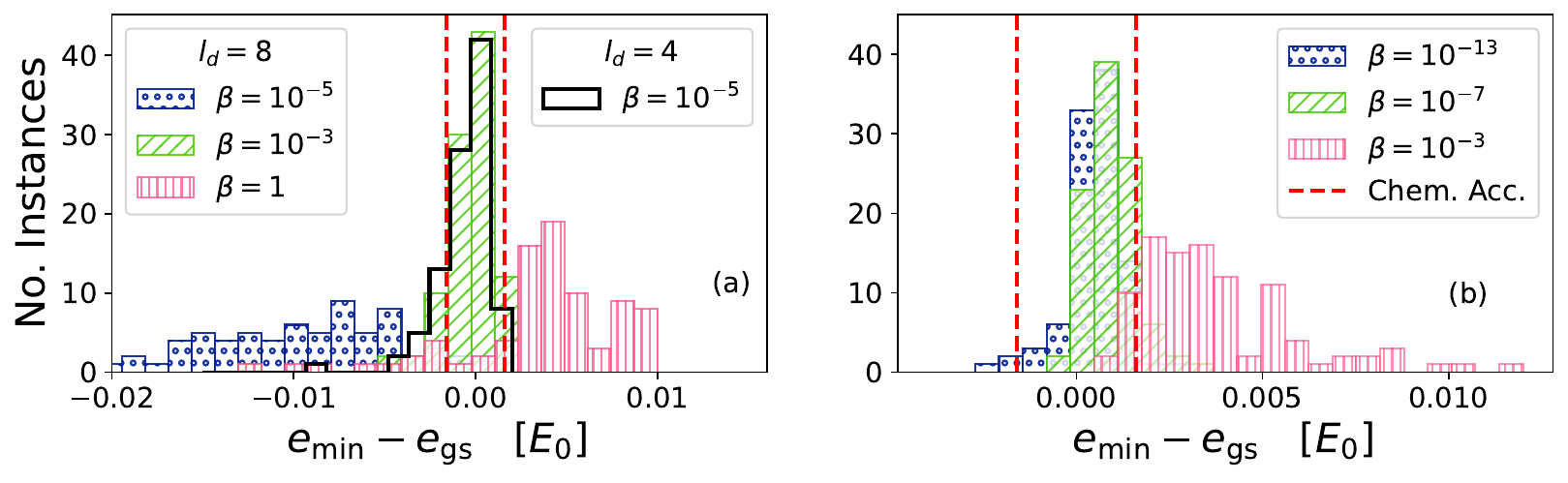}
\caption{\label{fig:histograms1d}Histograms of the relative energy discrepancies $e_{\min}-e_{\gs}$ for a test set of 1D Gaussian potentials (a) and 1D speckle potentials (b), after $t_{\max} = 15000$ steps of the gradient descent optimization. In (a), for $\beta=1$, the excessive  regularization of the latent space generates positive deviations. On the other hand, for $\beta=10^{-5}$ and $l_d=8$, substantial violations of variational property occur (blue bins with dot pattern). A more appropriate latent space dimension, $l_d=4$ for the 1D Gaussian model solves the issue (empty black bins). For the 1D speckle case (b), reducing $\beta$ and adopting the latent dimension $l_d=16$ improves the accuracy without affecting the stability. The energy unit $E_0=\hbar^2/ma_0^2$ can be identified with the Hartree energy. The vertical (red) lines indicate the chemical accuracy.}
\end{figure*}
\begin{figure*}
\includegraphics[width=\textwidth]{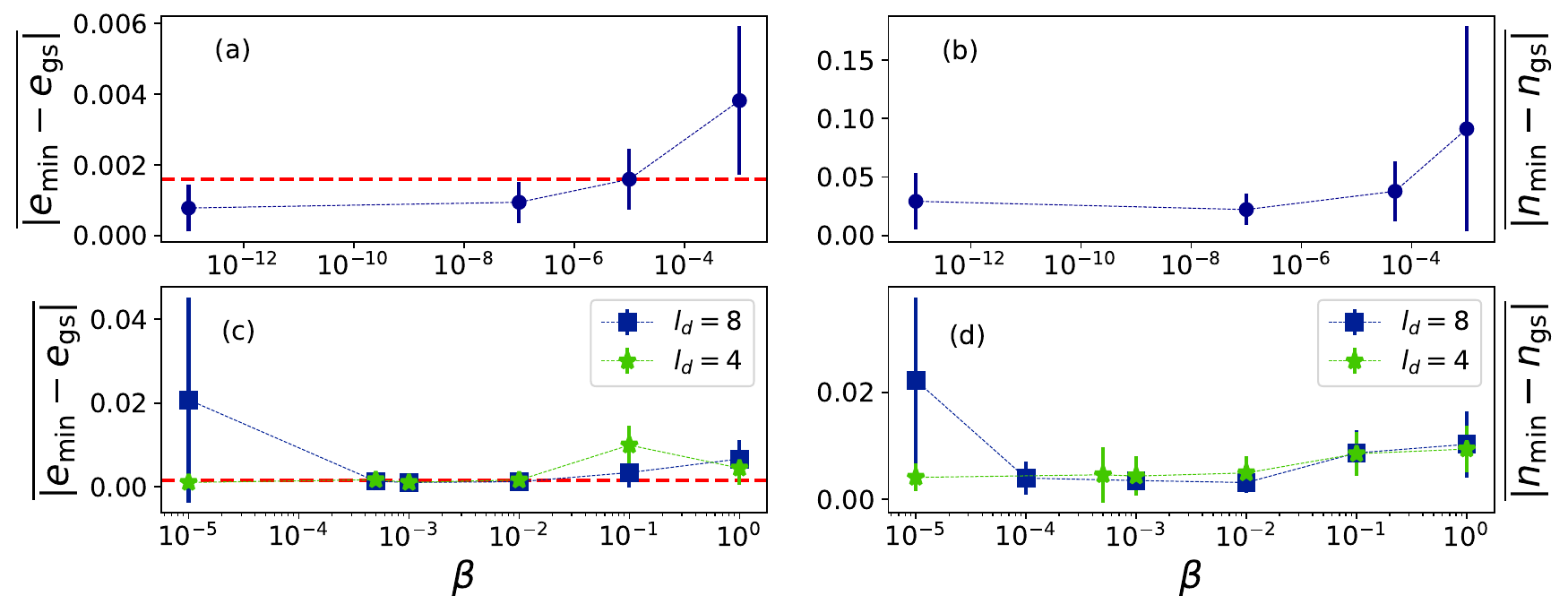}
\caption{\label{fig:4panels-vs-beta}Mean energy error $\overline{|e_{\min}-e_{\gs}|}$ [(a) and (c)] and mean reconstruction error $\overline{|\bm{n}_{\min}-\bm{n}_{\gs}|}$ [ (b) and (d)] after gradient descent optimization for 1D speckle potential [(a) and (b)] and the 1D Gaussian model [(c) and (d)].}
\end{figure*}
\begin{figure*}
\includegraphics[width=\textwidth]{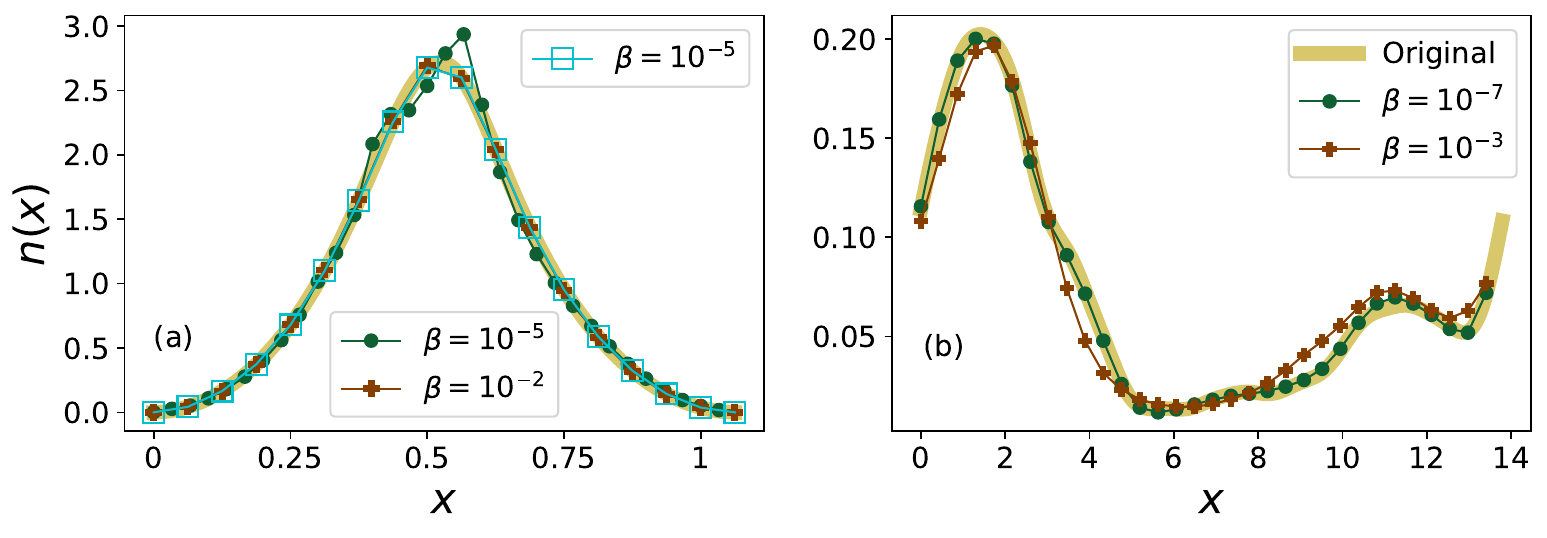}
\caption{\label{fig:samples-after-grad-desc}Illustrative examples of density profiles $n(x)$ obtained by minimizing the DL functional via gradient descent, for the 1D Gaussian potential (a) and for the 1D speckle potential (b). Different values of the regularization parameter $\beta$ are considered.}
\end{figure*}

\subsection{Density profile reconstruction by VAE}
The first analysis we perform aims at attesting only the reconstruction accuracy of the VAE. Specifically, we check whether the trained VAE is able to accurately replicate in output the density profiles provided in input. %Moreover, this means that the NN restricts effectively the density profile manifold. 
If successful, this test would demonstrate that VAEs allow creating  compressed representations of density profiles without loss of any relevant information.
As illustrative examples, for this preliminary test we mostly focus on the two 1D potentials.
Special attention is devoted to the role of the hyperparameter $\beta$, which tunes the relative weight attributed to the regularization loss compared to the reconstruction term.
The latent space dimension is set at the intermediate values $l_d=8$ and $l_d=16$ for the Gaussian and the speckle potentials, respectively. The (larger) choice for the latter is due to the enhanced variability of the corresponding density profiles, compared to those corresponding to the Gaussian model. The role of $l_d$ is further discussed in Sec.~\ref{subsec:grad-desc}.

The reconstruction is performed via the combined application of the encoder and the decoder on an input density profile:
\begin{equation}
    \hat{\bm{n}}= D_{\theta}[\bm{ \mu}_{\phi}[\bm{n}]].
\end{equation}
To quantify the reconstruction accuracy, we determine the average of the integrated absolute density difference, which we denote with $\overline{|\hat{\bm{n}}-\bm{n}|}$, and the  average is computed over a test set of $N_{\ts}=100$ samples.
As shown in Fig.~\ref{fig:reconstruction-vs-beta}, for both testbeds the reconstruction error rapidly decreases with $\beta$, allowing reaching faithful profiles in regimes where, as fully discussed in Sec.~\ref{subsec:grad-desc} in the framework of gradient-descent optimization, the latent space is still sufficiently regular.
It is worth mentioning already here that, if the VAE is not able to faithfully produce all ground-state densities, it is not  adequate to guide the gradient-descent minimization of the DL functional, since it might prevent reaching the actual ground state of some Hamiltonian instances.

A visual representation of the reconstruction accuracy for different values of the regularization parameter $\beta$ is provided in Fig.~\ref{fig:reconstruction-samples}; there, one representative Hamiltonian instance for each 1D testbed is considered.
One notices that large values of $\beta$ cause substantial distortions in the reconstructed profiles, which indeed do not  precisely reproduce all details of the ground truth data. 
Analogous findings are obtained for the 3D Gaussian potential. For example, for $\beta=10^{-6}$ and latent space dimension $l_d=32$, the reconstruction error is as small as $\overline{|\bm{n}-\hat{\bm{n}}|}=0.003(1)$.

\subsection{Gradient Descent Results} \label{subsec:grad-desc}

Hereafter, the VAE trained to encode ground-state density profiles (see Sec.~\ref{sec:results}) is employed to guide the energy minimization of the DL functional, following the approach described in Sec.~\ref{sec:vae-dft}.
Specifically, we analyze the accuracy of the gradient descent algorithm, inspecting whether the actual ground-state is reached. The goal is to avoid both spurious constraints that would lead to a positive bias, as well as instabilities due to unphysical profiles which often lead to negative biases, i.e., to violations of the variational property. The roles of the regularization parameter $\beta$ and of the latent space dimension $l_d$ are analyzed.

The first testbeds we discuss are the 1D Gaussian and speckle potentials.  Fig.~\ref{fig:histograms1d} displays histograms of energy discrepancies after gradient descent for both the 1D Gaussian case and the 1D speckle case.
When the latent space dimension is overestimated ($l_d=8$ for the Gaussian model), we observe both positive discrepancies in the over-regularized regime (large $\beta$) as well as sizable variational violations in the opposite regime (small $\beta$). These effects are more quantitatively analyzed in Fig.~\ref{fig:4panels-vs-beta}, where the average absolute discrepancies are shown as a function of $\beta$. Furthermore, illustrative instances of potentials and density profiles are shown in Fig.~\ref{fig:samples-after-grad-desc}.
One notices that excessively large values of $\beta$ do not allow constructing all details of the density profile, in particular for the speckle potential. This leads to positive energy discrepancies. Too small values introduce numerical artifacts, in particular for the Gaussian model, which cause the DL-functional to provide erroneous outputs and, hence, lead to negative energy discrepancies after the energy minimization. Yet, it relatively easy to tune $\beta$ and $l_d$ within a very broad range where the energy discrepancies are typically well below the threshold of chemical accuracy, generally set at 1kcal/mol. For example, with $l_d=4$ for the Gaussian model and $l_d=16$ for the (more variegate) speckle potential, $\beta$ can be tuned almost at will.

\begin{figure}
\includegraphics[width=\columnwidth]{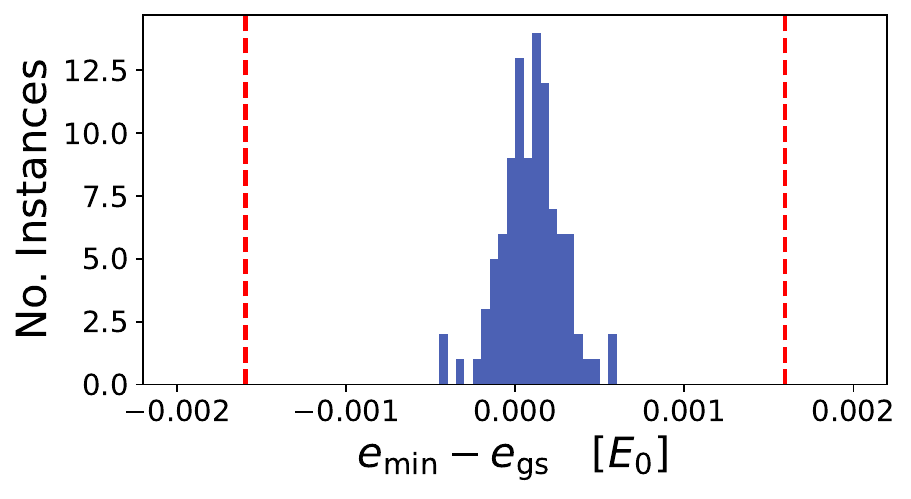}
\caption{\label{3D histogram}Histograms of the relative energy discrepancies $e_{\min}-e_{\gs}$ for a test set of 3D Gaussian potentials after gradient-descent optimization. The initial profiles are generated by VAE with $\beta=10^{-6}$. The vertical (red) lines indicate the threshold of chemical accuracy.}
\end{figure}

\begin{figure*}
\includegraphics[width=\textwidth]{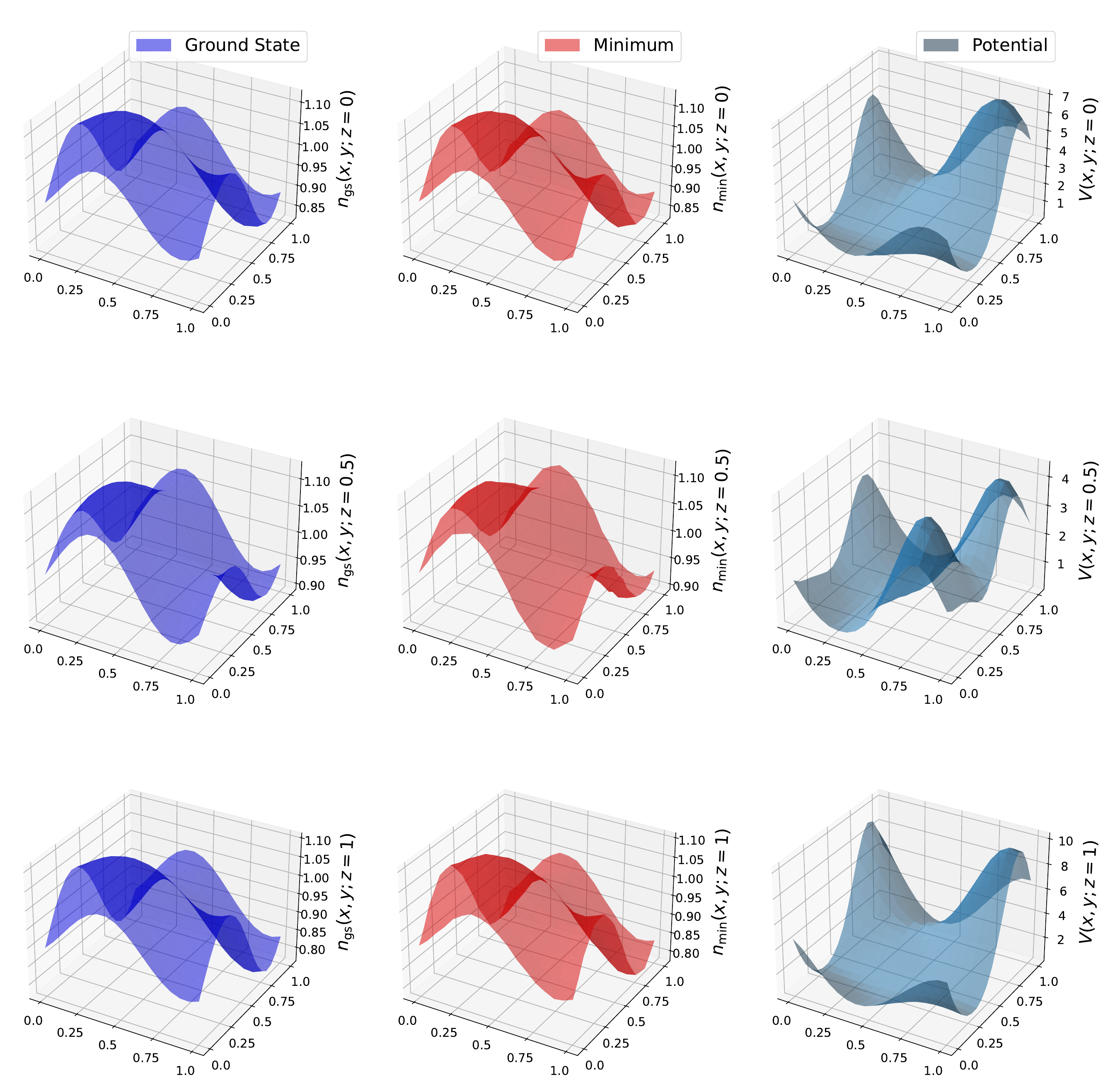}
\caption{\label{3D instance} An illustrative example of minimum density profile $n_{\min}(x,y,z)$ (red surfaces in the middle column) at different slices, compared to the corresponding ground state density profile $n_{\gs}(x,y,z)$ (blue surface, left column). The corresponding slices of the 3D Gaussian potential $V(x,y,z)$  are also shown (right column, unit of $E_0$).}
\end{figure*}

While low-dimensional testbeds for DL-DFT methods have been intensively investigated also in previous literature, here we extend our analysis to a challenging 3D potential, namely, the Gaussian-scatterer potential defined in Sec.~\ref{subsec:gaussian3D}.
In Fig.~\ref{3D histogram}, the energy discrepancies after the gradient descent are visualized, considering the hyperparameters $\beta=10^{-6}$ with $l_d=32$. 
Remarkably, the average absolute error is as small as $\overline{|e_{\min}-e_{\gs}|}=0.0002(2) \; E_0$.
Notice that, again, this is below the chemical accuracy.
To visualize the fidelity also of the reconstructed density profile, we show in Fig.~\ref{3D instance} three slices at different values of the $z$ coordinate. 
The dimensionless average density error after gradient descent is  $\overline{|\bm{{n}}_{\min}-\bm{n}_{\gs}|}=0.004(1)$, indicating that the density profiles are accurately reproduced also in this 3D testbed.

\section{Conclusions}
\label{sec:conclusion}

In this investigation, we have demonstrated that $\beta$-VAEs are suitable to create compressed and regular representations of the density profiles of rather variegate physical systems. 
Chiefly, the trained VAEs allowed us to implement an efficient and stable energy minimization of DL energy functionals, thus solving a critical problem in the growing field of DL-based DFT. 
Our strategy is based on the combination of two deep NNs, namely, the VAEs and a separate convolutional model that maps the density profile to the corresponding energy. Automatic algorithmic differentiation is adopted on the combined model, making use of this enabling feature provided by modern DL software. In turn, this also allowed us to efficiently perform the gradient-descent optimization, also exploiting the computational performance of graphic processing unit.
Our numerical analysis focused on three testbed models, including both Hamiltonians that, for different random parameters, lead to rather consistent density profiles, as well as the opposite case where density profiles vary substantially. Notably, we also addressed a 3D Hamiltonian.
The role of the regularization parameter $\beta$ and of the latent space dimension has been analyzed, showing that suitable parameters that avoid both violations of the variational property and positive variational biases can be easily identified.
To favor further studies on DL-DFT in 3D, we provide  at the repository of Ref.~\cite{emanuele_2024_10814856} a database suitable for training and testing deep NNs for the 3D Gaussian model.

DL approaches are being increasingly adopted with different goals in the framework of DFT (see, e.g., Refs.~\cite{Brockherde2017,10.1088/2516-1075/ac572f,10.1063/1.5025668,ALGHADEER2021127621,10.1063/5.0166432,FUJINAMI2020137358,dick2020machine,PhysRevLett.126.036401,li2022deep,Yang_2023,Zhang_2024,PhysRevA.101.050501,Yang_2023}). They are mostly used to learn energy-density functionals from data, but also to accelerate the implementation and the solution of DFT with conventional approximations (see, e.g., Refs.~\cite{doi:10.1021/acs.jpca.2c05922,10.1063/5.0138429}). For example, in a recent preprint~\cite{decamargo2023orbitalfree} autonormalizing flows have been used to sample the density profiles, allowing the minimization of a conventional orbital-free functional via Monte Carlo sampling. In our study, we combined the VAE with a DL-functional trained from data, avoiding Monte Carlo integration by decoding density profiles from the latent space.
From a general perspective, our study highlights the use of VAEs as a computational tool to extract effective variables that describe complex quantum systems in a compressed but essentially lossless manner.

\begin{acknowledgments}

We acknowledge useful discussions with R. Fazio, S. Cantori, and L. Brodoloni.
This work was supported by the PNRR MUR Project No. PE0000023-NQSTI and by the Italian Ministry of University and Research under the PRIN2022 project ``Hybrid algorithms for quantum simulators'' No. 2022H77XB7.
S.P. acknowledges support from the CINECA awards IsCa6\_NEMCAQS and IsCb2\_NEMCASRA, for the availability of high-performance computing resources and support.
We also acknowledge the EuroHPC Joint Undertaking for awarding this project access to the EuroHPC supercomputer LUMI, hosted by CSC (Finland) and the LUMI consortium through a EuroHPC Regular Access call.
\end{acknowledgments}

\bibliography{biblio}% Produces the bibliography via BibTeX.

\end{document}